\documentclass[10pt]{article}

\usepackage{amsmath}
\usepackage{amssymb}

\usepackage{graphicx}

\usepackage{cite}

\usepackage{color} 

\usepackage[labelfont=bf,labelsep=period,justification=raggedright]{caption}

\makeatletter
\renewcommand{\@biblabel}[1]{\quad#1.}
\makeatother

\date{}

\begin{document}

\begin{flushleft}
{\Large
\textbf{A maximum likelihood based technique for validating detrended fluctuation analysis (ML-DFA)}
}
\\
Maria Botcharova$^{1,2}$, 
Simon F. Farmer$^{2,\dagger}$, 
Luc Berthouze$^{3,4,\ast}$
\\
\bf{1} CoMPLEX,Centre for Mathematics and Physics in the Life Sciences and Experimental Biology/University College London, London, UK
\\
\bf{2} Institute of Neurology/University College London, London, UK
\\
\bf{3} Centre for Computational Neuroscience and Robotics/University of Sussex, Falmer, UK
\\
\bf{3} Institute of Child Health/University College London, London, UK
\\
E-mail: $\ast$ L.Berthouze@sussex.ac.uk, $\dagger$ s.farmer@ucl.ac.uk

\end{flushleft}

\section*{Abstract}
Detrended Fluctuation Analysis (DFA) is widely used to assess the presence of long-range temporal correlations in time series. Signals with long-range temporal correlations are typically defined as having a power law decay in their autocorrelation function. The output of DFA is an exponent, which is the slope obtained by linear regression of a log-log fluctuation plot against window size. However, if this fluctuation plot is not linear, then the underlying signal is not self-similar, and the exponent has no meaning. There is currently no method for assessing the linearity of a DFA fluctuation plot. Here we present such a technique, called ML-DFA. We scale the DFA fluctuation plot to construct a likelihood function for a set of alternative models including polynomial, root, exponential, logarithmic and spline functions. We use this likelihood function to determine the maximum likelihood and thus to calculate values of the Akaike and Bayesian information criteria, which identify the best fit model when the number of parameters involved is taken into account and over-fitting is penalised. This ensures that, of the models that fit well, the least complicated is selected as the best fit. We apply ML-DFA to synthetic data from FARIMA processes and sine curves with DFA fluctuation plots whose form has been analytically determined, and to experimentally collected neurophysiological data. ML-DFA assesses whether the hypothesis of a linear fluctuation plot should be rejected, and thus whether the exponent can be considered meaningful. We argue that ML-DFA is essential to obtaining trustworthy results from DFA. 

\section*{Introduction}

Detrended Fluctuation Analysis (DFA) is a technique commonly applied to time series as a means of approximating the Hurst exponent, which indicates the degree of long-range temporal correlations present~\cite{peng94,peng95,hurst,clegg}. Long-range temporal correlations (LRTCs) occur in time series with an autocorrelation function that decays as a power law function of the lag~\cite{granger}. The presence of LRTCs suggests that the underlying signal is governed by non-local behaviour, with all scales contributing to system behaviour. LRTCs have been detected in various biological time series and natural phenomena~\cite{hurst,peng94,peng95,varotsos,robinson,karmeshu,karagiannis}, see a review in~\cite{samorodnitsky}. In neurophysiological signals, it has been argued that LRTCs facilitate essential functions such as memory formation, rapid information transfer, and the efficient neural network reorganisation that promotes learning~\cite{chialvo,sornette,timme,stam,werner,linkenkaer01,linkenkaer04}.

DFA produces estimates of the magnitude of detrended fluctuations at different scales (window sizes) of a time series and assesses the scaling relationship between estimates and time scales. Estimation of the Hurst exponent through DFA assumes self-similarity in the time series. If the signal is self-similar, then the detrended fluctuations will increase as a power law function of window size, and the relationship between the two can be visualised as a straight line on a log-log fluctuation plot \cite{peng94,peng95}. DFA returns the slope of the plot as its exponent with no check as to whether the self-similarity of the time series is supported by there being a linear fluctuation plot. At present there is no method which establishes the linearity of a DFA plot and an important shortcoming of the typically used method (see below) is that unless there is gross violation of linearity that can be detected by visual inspection then DFA exponents can be used for data that are not self-similar. In fact, Maraun and colleagues went as far as suggesting that DFA results are sensitive but not specific concerning long-range correlations~\cite{maraun04}.  

Previous studies have described non-linear characteristics in DFA fluctuation plots for signals constructed by independent superposition of a number of processes with specific characteristics. When a noise time series contains a linear, sinusoidal, or a power law trend, the DFA plot will contain several linear segments, joined at crossover points~\cite{hu}. Studies have also looked at noise time series with sections of silence, concatenations of noisy signals with different amplitude standard deviations, and of noisy signals with varying levels of temporal correlation~\cite{chen02,chen05}. These fluctuation plots show different combinations of linear and quasi-linear fragments. 

At present the standard approach used to characterise the fit of the linear regression is to calculate an $R^2$ value (for example~\cite{grech}). However, the $R^2$ value is a very insensitive measure~\cite{anscombe}. An alternative technique may be to assume that the errors around a linear fit have a $\chi^2$ distribution, but this assumption cannot be made for a DFA fluctuation plot because the magnitude of detrended fluctuation is dependent on the window length so that the fluctuation plot suffers from heteroscedasticity~\cite{raymond}. Namely, this approach would not allow one to distinguish between a self-similar signal yielding non $\chi^2$-distributed regression errors and a non self-similar signal. Another approach may be to compute the probability of the fluctuation plot taking the form of a specific function for signals with different self-similarity properties, based on the probability distribution of the innovations. A paper by Bardet~\cite{bardet} formulates such a distribution for the scale-invariant process that would give rise to a perfectly linear plot in its DFA fluctuation plot, i.e., fractional Gaussian noise. However, this approach would restrict the technique to being able to identify only a limited set of signals, and furthermore, one would need to know {\it a priori} the nature of the signal in order to employ the appropriate distribution. 

Here, we propose a maximum likelihood based technique to assess the validity of the assumption of linearity through model selection. 

Our technique, referred to as ML-DFA henceforth, is rooted in likelihood theory. We calculate a log-likelihood function for both a linear model and a number of alternative models. This requires formulating the DFA fluctuation plot as a probability density, which we do by normalising the fluctuation magnitudes. We use this to compute the Akaike and Bayesian Information Criterion (AIC and BIC, respectively)~\cite{akaike,schwartz} which reveal the best-fitting model to the fluctuation plot, while compensating for over-fitting. If no model amongst the set of alternative models is a better fit than the linear model, then we accept (or more accurately, we do not reject) the hypothesis that the fluctuation plot is linear.  

In the following sections, we apply the method to simulated time series for which we can control the expected outcome, and to neurophysiological data for which no ground truth is available. 

Synthetic time series are generated by an Autoregressive Fractionally Integrated Moving Average (FARIMA) process~\cite{hosking} (also referred to as ARFIMA or AFRIMA). We use FARIMA because it provides an easily tunable algorithm for constructing time series with a combination of short-term and long-term correlations, which we will show influence the DFA fluctuation plot. A FARIMA process in its simplest form can be used to generate fractional Gaussian noise, which has been shown analytically to produce linear DFA fluctuation plots in their asymptotic limit~\cite{taqqu,bardet}. However, by gradually introducing short-term correlations through smoothing the data and enforcing autoregression, it is possible to destroy the self-similarity of the time series, and a statistically robust method should capture this. We note that a FARIMA process has also been used to model neurophysiological signals such as EEG, which have the properties of being stationary and whose amplitude fluctuations follow Gaussian statistics~\cite{mcewen}. A FARIMA process therefore provides an efficient and malleable method of generating and manipulating time series. We further apply ML-DFA to a sinusoidal signal and a sinusoidal signal with independent additive noise, whose DFA fluctuation plots take known forms~\cite{hu}. 

Finally, we apply the method to EEG data recorded from a group of twenty human subjects at rest to demonstrate the technique on experimentally acquired data. We also study how the choice of window lengths over which DFA is calculated affects the linearity or otherwise of the fluctuation plot.

\section*{Results}\label{res}

\subsection*{Simulated Data}

We applied ML-DFA to simulations of a FARIMA process. We first used FARIMA to generate self-similar fractional Gaussian noise with varying Hurst exponents, and then we altered its parameters to generate a more general non self-similar FARIMA signal. For a more complete discussion of the FARIMA process parameters, the reader is referred to the Methods and Materials section below. We further applied ML-DFA to sinusoidal signals with three different periods, and to sinusoidal signals with independently added noise.

We fitted the DFA fluctuation plots for $1000$ simulations of each of the generated time series with the set of alternative models listed in the Methods and Materials section. We report the proportion of best-fits for each model as determined by the AIC and BIC measures. 

\subsubsection*{Fractional Gaussian Noise}

Fractional Gaussian noise can be generated by a FARIMA(0,$d$,0) process with $0<d<0.5$. The case of $d=0$ is called white Gaussian noise, however, we will here refer collectively to FARIMA(0,$d$,0) processes with $0 \leq d<0.5$ as fractional Gaussian noise. Fractional Gaussian noise has been proved to be asymptotically scale-invariant, and therefore its associated DFA fluctuation plot should be linear with a slope $\alpha$ given by $d+0.5$ \cite{taqqu,bardet}. The value of $\alpha$ is an approximation to the Hurst exponent of the data, $H$, where $H=0.5$ indicates Gaussian white noise and $H=1$ indicates pink noise. We demonstrate ML-DFA on three simulations of fractional Gaussian noise, spanning the possible range of $d$ values.

\begin{figure}[!ht]
\begin{center}
\includegraphics[scale=0.5]{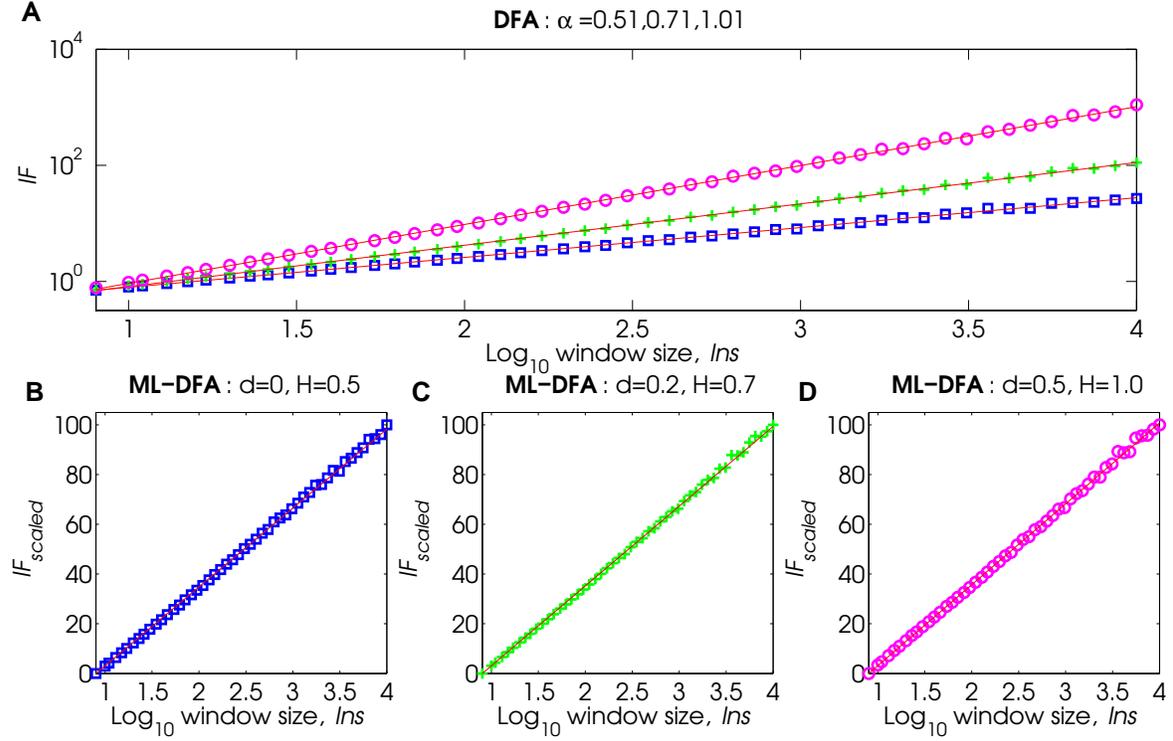}
\end{center}
\caption{{\bf Time series and corresponding DFA fluctuation plots for signals obtained by FARIMA(0,$d$,0) processes with $\phi$ and $\theta$ set to $0$ and with $d = 0$, and $d=0.2$, and $d=0.5$ to produce fractional Gaussian noise.} Panel A shows the three DFA fluctuation plots fitted using standard DFA. Values $d = 0$, $d=0.2$, and $d=0.5$ will produce time series with Hurst exponents $0.5$ (white noise, blue diamonds), $0.7$ (correlated noise, green crosses) and $1$ (pink noise, pink circles) respectively. The slopes estimated by application of standard DFA are stated at the top, and correspond closely to these theoretical values. Panels B-D show the best fit model according to the AIC measure in ML-DFA. The best-fit model is linear in all cases.} \label{Fig1}
\end{figure}

In Figure~\ref{Fig1}A we show DFA plots for three FARIMA time series with Hurst exponents of $0.5$, $0.7$ and $1.0$. The slopes of the DFA plots recover estimates of the Hurst exponents of $0.50$, $0.71$ and $1.01$ respectively. Figure~\ref{Fig1}B-D shows that the results of ML-DFA confirm that a linear model is appropriate for each of the time series, thus validating the results of standard DFA.

Tables~\ref{bic} and~\ref{aic} report the proportion of times out of $1000$ simulations that each of the alternative models was found by ML-DFA to be the best fit, according to the AIC and BIC values respectively. We found that the AIC and BIC were both successful at identifying the linear model as the best fit in over $95$\% of the simulations. The mean slopes of those fluctuation plots that were not rejected were the same to 3 decimal places for both the AIC and BIC measures, and were $0.500$, $0.696$ and $0.995$ respectively for expected Hurst parameters of $0.5$, $0.7$ and $1.0$. The standard deviations for all slopes were 0.01.  

\subsubsection*{FARIMA processes}

The FARIMA($1$,$d$,$1$) process is one which includes a single $\phi$ and a single $\theta$ coefficient, indicated by the parameter values of $1$. It is possible to include a greater number of $\phi$ and $\theta$ coefficients, but we consider only a single addition for simplicity. We vary $\phi$ and $\theta$ in the range $0<\phi<1$, $0<\theta<1$, which satisfies the conditions $\vert \phi \vert < 1$, $\vert \theta \vert < 1$ for convergence \cite{hosking}. Throughout the manuscript, FARIMA([$\phi$],$d$,[$\theta$]) will denote the FARIMA process with $\phi_1=\phi$ and $\theta_1=\theta$. 

\begin{figure}[!ht]
\begin{center}
\includegraphics[scale=0.5]{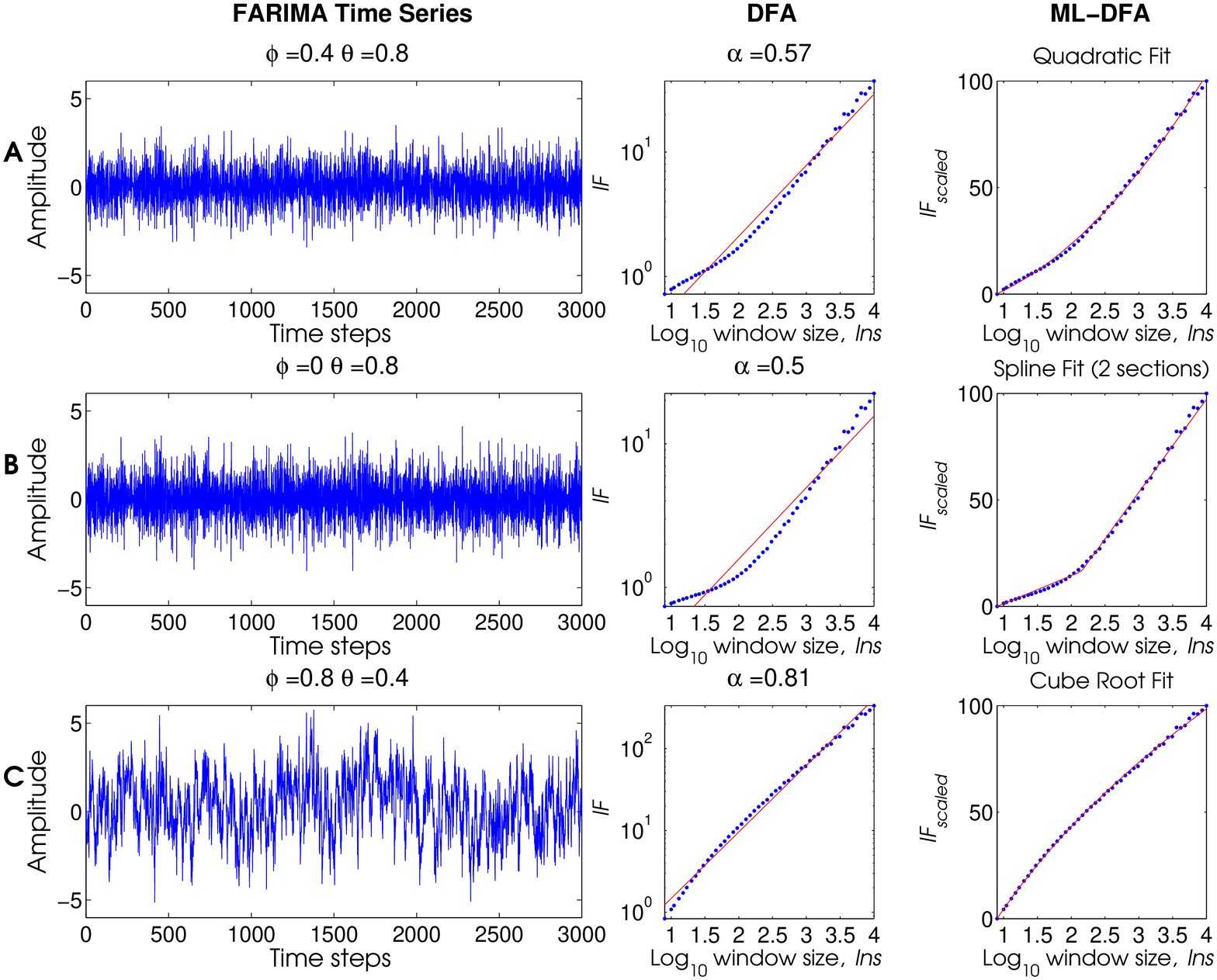}
\end{center}
\caption{{\bf Time series and corresponding DFA fluctuation plots for signals obtained by FARIMA(1,0.2,1) processes with $d=0.2$ taken as an representative value, and varying values of $\phi$ and $\theta$.} Each row A-C corresponds to a different set of $\phi$ and $\theta$ coefficients, which alter the resulting DFA fluctuation plots. In each row, the left-hand side panel shows a representative $3000$ innovations of the time series, the middle panel shows the fluctuation plot fitted using standard DFA with the resulting exponent $\alpha$ given above, and the right-hand side panel shows the best-fit model determined by ML-DFA using AIC.} \label{Fig2}
\end{figure}

In the general case, a FARIMA($1$,$d$,$1$) time series is not expected to be self-similar and therefore, the associated DFA fluctuation plots should not necessarily be linear. Variations in the $\phi$ and $\theta$ parameters contribute to a range of fluctuation plots, with examples in Figure \ref{Fig2} illustrating a number of cases in which different alternative models were found by ML-DFA to be the best fit.  

\begin{figure}[!ht]
\begin{center}
\includegraphics[scale=0.5]{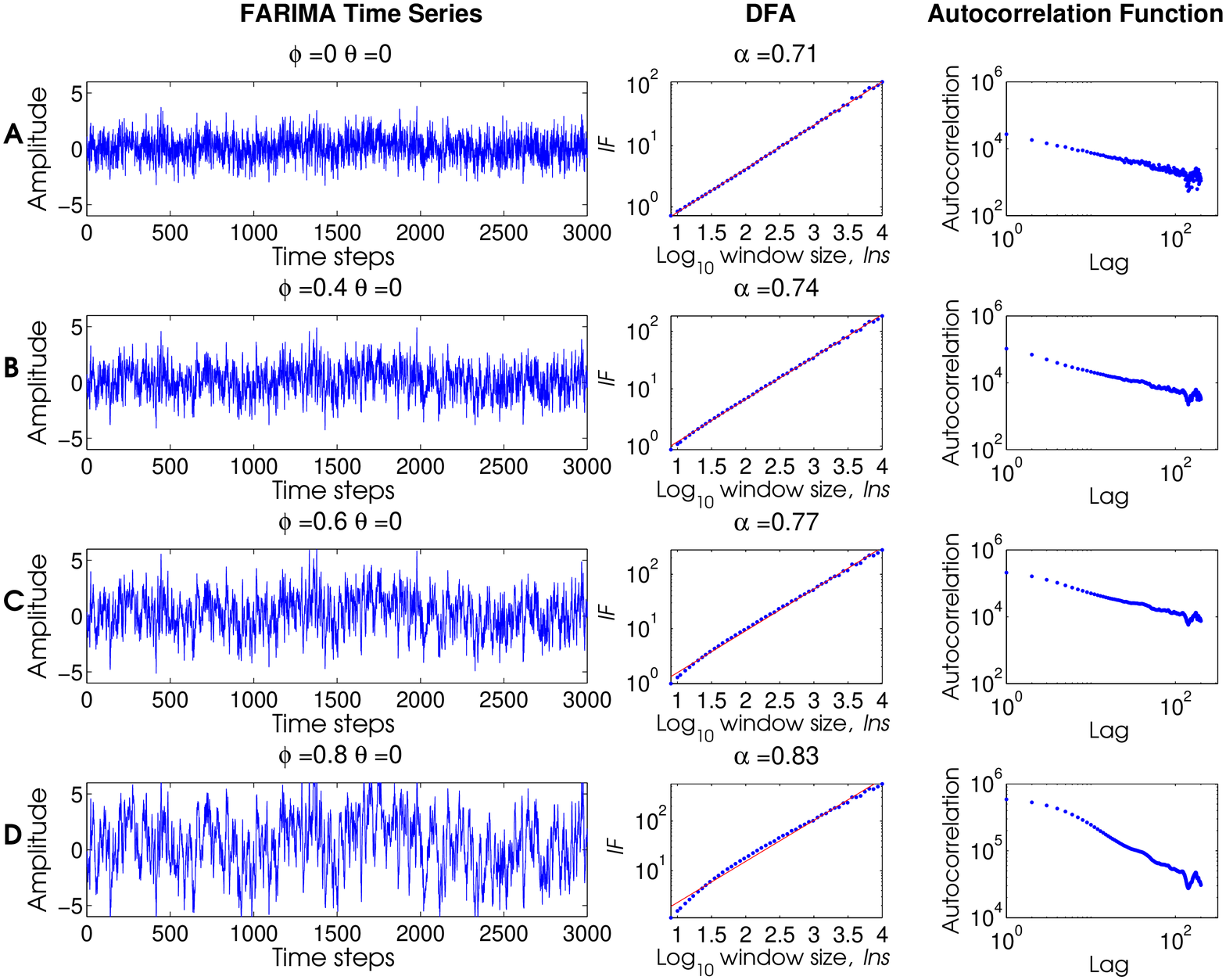}
\end{center}
\caption{{\bf Time series, DFA fluctuation plots and autocorrelation functions for signals obtained by FARIMA([p],0.2,[0]) processes with varying values of $p$.} Each row A-D corresponds to $\phi$ taking values from ${0,0.4,0.6,0.8}$.  In each row, the left-hand side panel shows a representative $3000$ innovations of the time series, the middle panel shows the fluctuation plot fitted using standard DFA with the resulting exponent $\alpha$ given above, and the right-hand side panel shows the autocorrelation function of the (complete) signal in log-log coordinates.} \label{Fig3}
\end{figure}

In all cases, ML-DFA showed sensitivity to more or less subtle deviations from the linear model. Tables~\ref{bic} and~\ref{aic} provide the proportion of times out of $1000$ simulations of FARIMA(1,$d$,1) time series that each of the alternative models were found by ML-DFA to be the best fit, using the AIC and BIC values respectively. In most cases (6/9 pure FARIMA scenarios), the results of AIC and BIC were compatible. Specifically, 4 of the 9 scenarios showed almost identical results (scenarios 1, 2, 6, 9) whilst 2 were qualitatively similar in that the same models were involved albeit with varying percentages (scenarios 3 and 8). Three scenarios showed substantial differences with different models being involved (scenarios 4, 5 and 7). However, on the most important point of whether the linear model hypothesis was to be rejected, there was strong agreement (8/9 pure FARIMA scenarios, 10/11 all scenarios) between AIC and BIC, the only exception being FARIMA([0.4],0.2,[0]) where BIC did not reject the linear model hypothesis in 31.1\% of the runs (to be contrasted with $2.1$\% for AIC). This once again illustrates that because the BIC is less likely to select models with a greater number of parameters, it may show less sensitivity to very subtle deviations from the expected model, thus showing more false positives than AIC. To illustrate this point, we re-examined the FARIMA([0.4],0.2,[0]) scenario systematically varying the value of the $\phi$ parameter. Figure~\ref{Fig3} shows that when $\phi=0.4$ the DFA fluctuation plot could certainly be considered linear on visual inspection. However, closer examination (we assist the reader by providing a log-log plot of the autocorrelation function -- it is helpful to remember that the DFA exponent is directly linked to the exponent of the power law in the autocorrelation function) reveals otherwise. Application of the runs test~\cite{runstest} on the residuals of the regression shows that the residuals are not independent ($p<1e-5$) which confirms that the BIC results are false positives. For comparison, the runs test for the fractional Gaussian noise returns $p>.2$. With increasing values of $\phi$, the distortion of the fluctuation plot (and associated autocorrelation function) becomes readily available to visual inspection and agreement between AIC and BIC is strong. Specifically, $100\%$ of the simulations reject the linear model hypothesis and AIC and BIC return the same set of alternative models in more than $99\%$ of the simulations for both $\phi=0.6$ and $\phi=0.8$.

\begin{table}[ht]
\caption{\bf{ML-DFA results on synthetic data using AIC. From 1000 simulations of noise time series, the table gives the proportion of times that each of the alternative models was found to be the best fit, according to AIC values when ML-DFA was applied to fractional Gaussian noise, FARIMA(1,$d$,1) processes and noisy sinusoidal signals.}}  
\centering 
\begin{tabular}{|l|c|c|c|c|c|c|c|c|c|c|} 
\hline 
& Linear & \multicolumn{9}{c|}{Non-linear} \\
\hline
Model &$x$& $x^2$ &$x^3$ &$x^4$&$\sqrt[n]{x}$ &4-$x$ & 3-$x$&2-$x$&log&$e$ \\
 [0.5ex] 
\hline 
FARIMA([0],0.5,[0])	&	96.6	&	3	&	-	&	-	&	-	&	-	&	-	&	-	&	-	&	0.4	\\
FARIMA([0],0.2,[0])	&	96.3	&	2.9	&	-	&	-	&	-	&	-	&	-	&	-	&	-	&	0.8	\\
FARIMA([0],0,[0])	&	95.9	&	2.8	&	-	&	-	&	0.1	&	-	&	-	&	-	&	0.1	&	1.1	\\
FARIMA([0.4],0.2,[0])	&	2.1	&	7.5	&	-	&	-	&	56.0	&	-	&	-	&	-	&	14.5	&	19.9	\\
FARIMA([0],0.2,[0.4])	&	-	&	77	&	19.3	&	0.1	&	-	&	-	&	-	&	1.2	&	-	&	2.4	\\
FARIMA([0.8],0.2,[0])	&	-	&	-	&	0.3	&	-	&	90.3	&	-	&	-	&	-	&	7.1	&	2.3	\\
FARIMA([0],0.2,[0.8])	&	-	&	47.7	&	2.8	&	2.8	&	-	&	-	&	-	&	45.5	&	-	&	1.2	\\
FARIMA([0.4],0.2,[0.8])	&	-	&	64.7	&	-	&	-	&	-	&	-	&	-	&	22.1	&	-	&	13.2	\\
FARIMA([0.8],0.2,[0.4])	&	-	&	0.4	&	-	&	-	&	76.5	&	-	&	-	&	-	&	13.5	&	9.6	\\
FARIMA([0],-0.2,[0])+sin($\frac{2 \pi t}{200}$)	&	-	&	-	&	-	&	-	&	-	&	7.4	&	92.6	&	-	&	-	&	-	\\
FARIMA([0],0,[0])+sin($\frac{2 \pi t}{100}$)	&	-	&	-	&	-	&	2	&	-	&	98	&	-	&	-	&	-	&	-	\\
\hline 
\end{tabular}
\begin{flushleft} The fitted models are listed in the top row, alongside the proportion of best fits assigned to each one by the value of the AIC measure. The shorthand n-$x$ is used to denote a n-segment spline. The shorthand $\sqrt[n]{x}$ combines results for $n=2,3,4$. The signals whose DFA fluctuation plots are analysed are described in the left-hand side column. 
\end{flushleft}
\label{aic} 
\end{table}

\begin{table}[ht]
\caption{\bf{ML-DFA results on synthetic data using BIC. From 1000 simulations, the table gives the proportion of times that each of the alternative models was found to be the best fit, according to BIC values when ML-DFA was applied to fractional Gaussian noise, FARIMA(1,$d$,1) processes and noisy sinusoidal signals.}}  
\centering 
\begin{tabular}{|l|c|c|c|c|c|c|c|c|c|c|} 
\hline 
& Linear & \multicolumn{9}{c|}{Non-linear} \\
\hline 
Model& $x$& $x^2$ & $x^3$ & $x^4$  & $\sqrt[n]{x}$ &4-$x$ & 3-$x$ &2-$x$ &log &$e$ \\
 [0.5ex] 
\hline 
FARIMA([0],0.5,[0])	&	96.6	&	3	&	-	&	-	&	-	&	-	&	-	&	-	&	-	&	0.4	\\
FARIMA([0],0.2,[0])	&	96.3	&	2.9	&	-	&	-	&	-	&	-	&	-	&	-	&	-	&	0.8	\\
FARIMA([0],0,[0])	&	96.8	&	2.3	&	-	&	-	&	0.1	&	-	&	-	&	-	&	0.1	&	0.7	\\
FARIMA([0.4],0.2,[0])	&	31.1	&	4.6	&	-	&	-	&	40.6	&	-	&	-	&	-	&	11.4	&	12.3	\\
FARIMA([0],0.2,[0.4])	&	0.1	&	93.9	&	1.6	&	-	&	-	&	-	&	-	&	0.5	&	-	&	3.9	\\
FARIMA([0.8],0.2,[0])	&	-	&	-	&	-	&	-	&	90.4	&	-	&	-	&	-	&	7.1	&	2.5	\\
FARIMA([0],0.2,[0.8])	&	-	&	71.2	&	1.7	&	-	&	-	&	-	&	-	&	23.6	&	-	&	3.5	\\
FARIMA([0.4],0.2,[0.8])	&	-	&	86.3	&	-	&	-	&	-	&	-	&	-	&	0.4	&	-	&	13.3	\\
FARIMA([0.8],0.2,[0.4])	&	-	&	0.4	&	-	&	-	&	76.5	&	-	&	-	&	-	&	13.5	&	9.6	\\
FARIMA([0],-0.2,[0])+sin($\frac{2 \pi t}{200}$)	&	-	&	-	&	-	&	-	&	-	&	7.4	&	92.6	&	-	&	-	&	-	\\
FARIMA([0],0,[0])+sin($\frac{2 \pi t}{100}$)	&	-	&	-	&	-	&	15.1	&	-	&	84.9	&	-	&	-	&	-	&	-	\\
\hline 
\end{tabular}
\begin{flushleft} The fitted models are listed in the top row, alongside the proportion of best fits assigned to each one by the value of the BIC measure. The shorthand n-$x$ is used to denote a n-segment spline. The shorthand $\sqrt[n]{x}$ combines results for $n=2,3,4$. The signals whose DFA fluctuation plots are analysed are described in the left-hand side column.   
\end{flushleft}
\label{bic} 
\end{table}

\subsubsection*{Sinusoidal signals}

\begin{figure}[!ht]
\begin{center}
\includegraphics[scale=0.5]{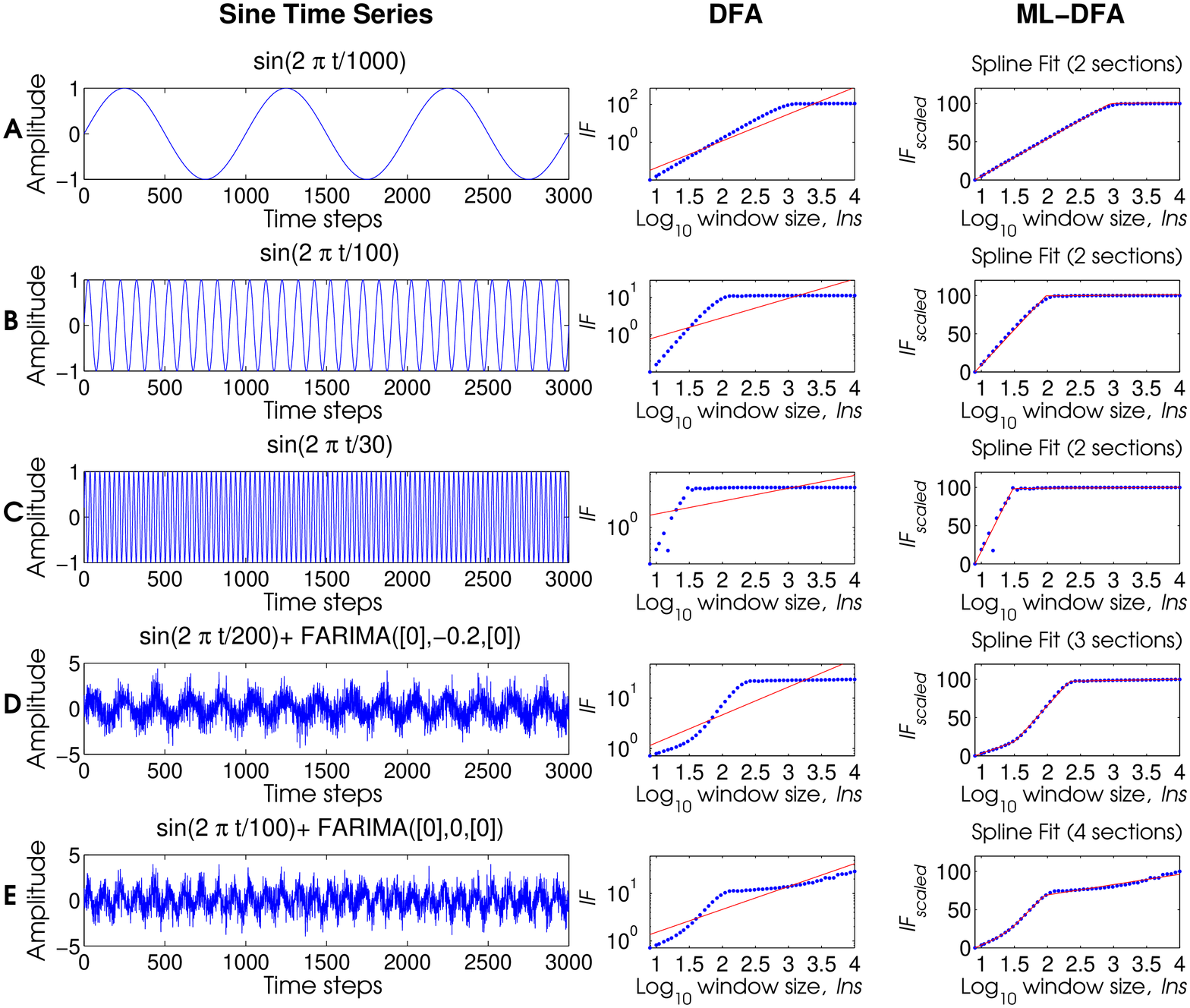}
\end{center}
\caption{{\bf Time series and corresponding DFA fluctuation plots for 5 sinusoidal signals with varying levels of independent, additive noise.} Each row A-E corresponds to a different sinusoidal function. In each row, the left-hand side panels show a representative $3000$ innovations of the time series, the middle panel shows the fluctuation plots fitted using standard DFA, and the right-hand side panel shows the best-fit model as determined by ML-DFA using AIC. \newline\newline} \label{Fig4}
\end{figure}

From \cite{hu}, the DFA fluctuation plot of a pure sine will have a crossover at a window size corresponding to the period of the oscillation, with a slope of $2$ for low window sizes, and a slope of zero after the crossover point. We reproduce these fluctuation plots and demonstrate that they are best fit by a two-segment spline model, with crossovers as predicted by theory. In Figure~\ref{Fig4}A-C, we present results for three pure sine curves with periods of $1000$, $100$ and $30$ respectively. We observe that the crossover points in each plot are at $3$, $2$ and $1.48$, which are the base-10 logarithms of $1000$, $100$ and $30$, respectively. ML-DFA therefore recovers both the spline function and at its point of inflection the period of the original sine signal.

The addition of independent noise to sinusoidal signals has also been studied~\cite{hu}. The DFA fluctuation plot of a sine signal with anti-correlated noise (Hurst exponent $H \in [0,0.5)$) will have two crossover points, and therefore three segments. One will be located at the window size corresponding to the period of the oscillation, and one at a smaller window length. We demonstrate in Figure~\ref{Fig4}D that ML-DFA identifies a three-segment spline as the best fitting model for such a fluctuation plot.  

A sine curve with independent, additive white or correlated noise will show three crossovers, or four segments in its DFA fluctuation plot. One crossover is again at the period of the sine curve. Figure~\ref{Fig4}E demonstrates ML-DFA alongside its resulting best-fit four-segment spline. 

Tables~\ref{bic} and~\ref{aic} provide the proportion of times out $1000$ simulations of two sets of sines with added noise that each of the alternative models was found to be the best fit by the AIC and BIC measures, respectively. No data are provided for repeated simulations of sines without added noise since these would produce rigorously identical fluctuation plots. Compared to the BIC measure, we found that the AIC measure assigned a greater proportion of the DFA fluctuation plots obtained from the sine with FARIMA([0],0,[0]) noise to the four-segment spline model ($98\%$ vs $84.9\%$ for AIC and BIC, respectively), as predicted by theory~\cite{hu}. The BIC measure returned a higher proportion of quartic model because of the reduced number of parameters. However, both AIC and BIC performed similarly in identifying the three-segment spline as the best fit for fluctuation plots of the sine with FARIMA([0],-0.2,[0]) noise, as expected.

\subsection*{Physiological Data}\label{physdat}

We applied ML-DFA to EEG data, according to the method set out in Linkenkaer-Hansen et al.~\cite{linkenkaer01}. Specifically, we took the power spectrum of the EEG, found the peak corresponding to the alpha rhythms and bandpass filtered the signal to isolate the corresponding range. Following this, we obtained the amplitude envelope by using the Hilbert transform, and applied standard DFA and ML-DFA. 

\begin{figure}[!ht]
\begin{center}
\includegraphics[scale=0.5]{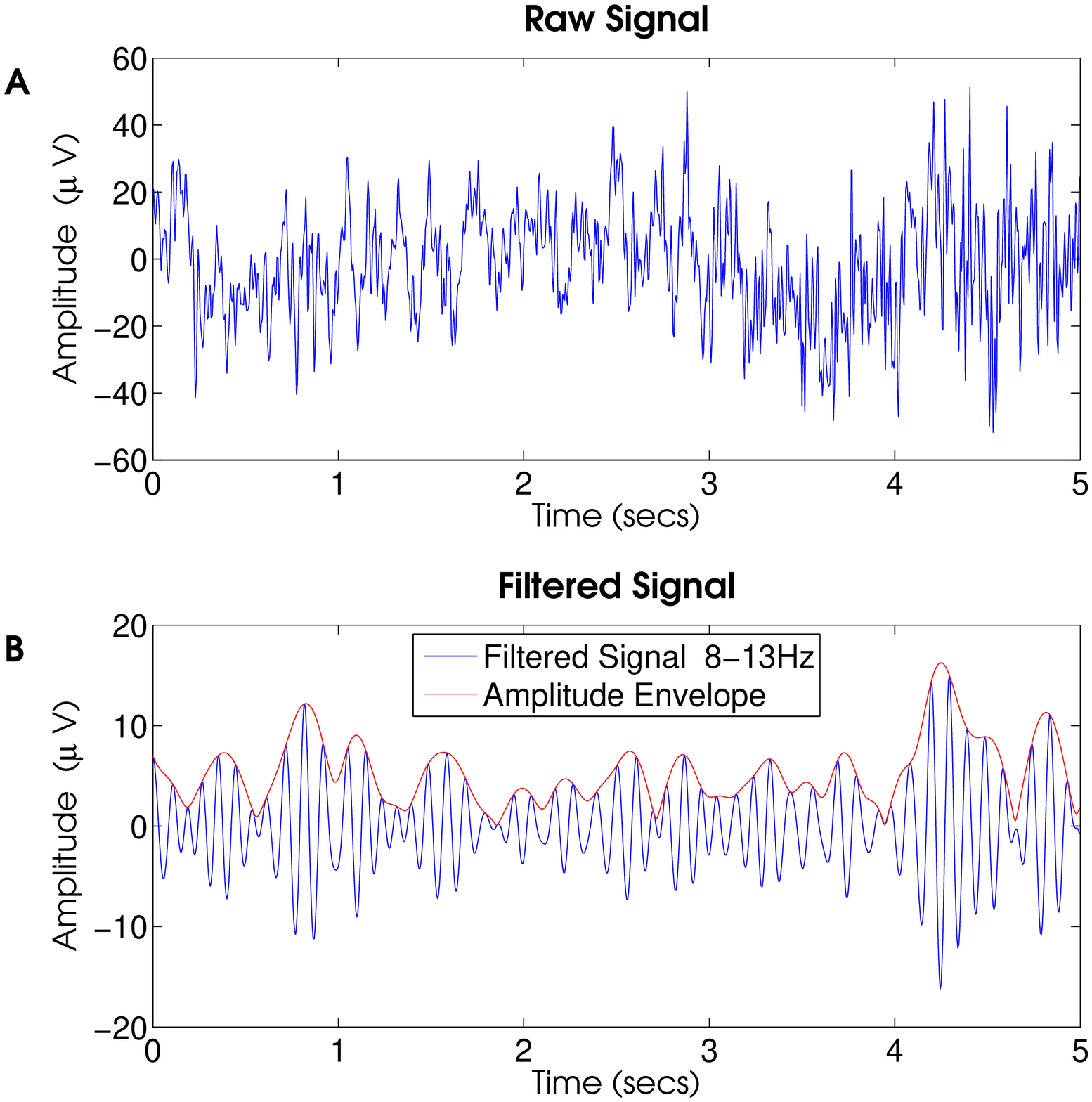}
\end{center}
\caption{{\bf Preprocessing of the time series on an example EEG data set.} Panel A shows the raw EEG signal. This is filtered between $8-13$ Hz for Panel B, and the amplitude envelope, derived from the real part of the Hilbert transform, is plotted above the filtered data.} \label{Fig5}
\end{figure}

The amplitude envelope was obtained by first applying the Hilbert transform to the time series $s(t)$ in order to obtain its analytic signal $ s_a$, which is a corresponding unique complex representation of a real-valued time series:
$$ s_a(t) = s(t) + H\left\lbrace s(t) \right\rbrace $$
where the Hilbert transform is represented by $H\left\lbrace  \right\rbrace $. The time-varying envelope $A(t)$ is then the amplitude of the analytic signal, given by:
$$ A(t) = \sqrt{s(t)^2 + H\left\lbrace s(t) \right\rbrace^2 }.$$
We demonstrate these steps in Figure \ref{Fig5}. 

DFA and ML-DFA were applied to $A(t)$, the amplitude envelope of an EEG time series filtered between 8 and 13 Hz. The minimum box size for applying DFA was $1$ second of data, in order to include at least 8 oscillations at the minimum frequency of the band-pass filter. The largest window size was set to one tenth of the full length of the data for each subject, as suggested by~\cite{linkenkaer01}. This allows a sufficient number of windows to provide a robust measure of the average fluctuation magnitude for a large window length, thus correcting for the variability of root mean square fluctuations from one window to the next. Note that, in~\cite{linkenkaer01}, the window sizes are determined by inspecting a fluctuation plot that spans across all possible window sizes, and then the range of windows that adhere to a power law is selected for further analysis. We will return to this in the Discussion.

\subsubsection*{Human EEG Data}
We report the best fit models as determined by ML-DFA for the amplitude envelope of the EEG of 20 human subjects tested, which had previously been filtered between 8 and 13 Hz. For each subject, an EEG time series from the Cz electrode was used after artefact removal because of its central location on the scalp, leading to fewer potential artefacts caused by muscle movements or eye-blinks. If the best fit model, as assessed by the BIC value, is linear, then we also report the DFA exponent in Table~\ref{healthyTab}. 

\begin{figure}[!ht]
\begin{center}
\includegraphics[scale=0.5]{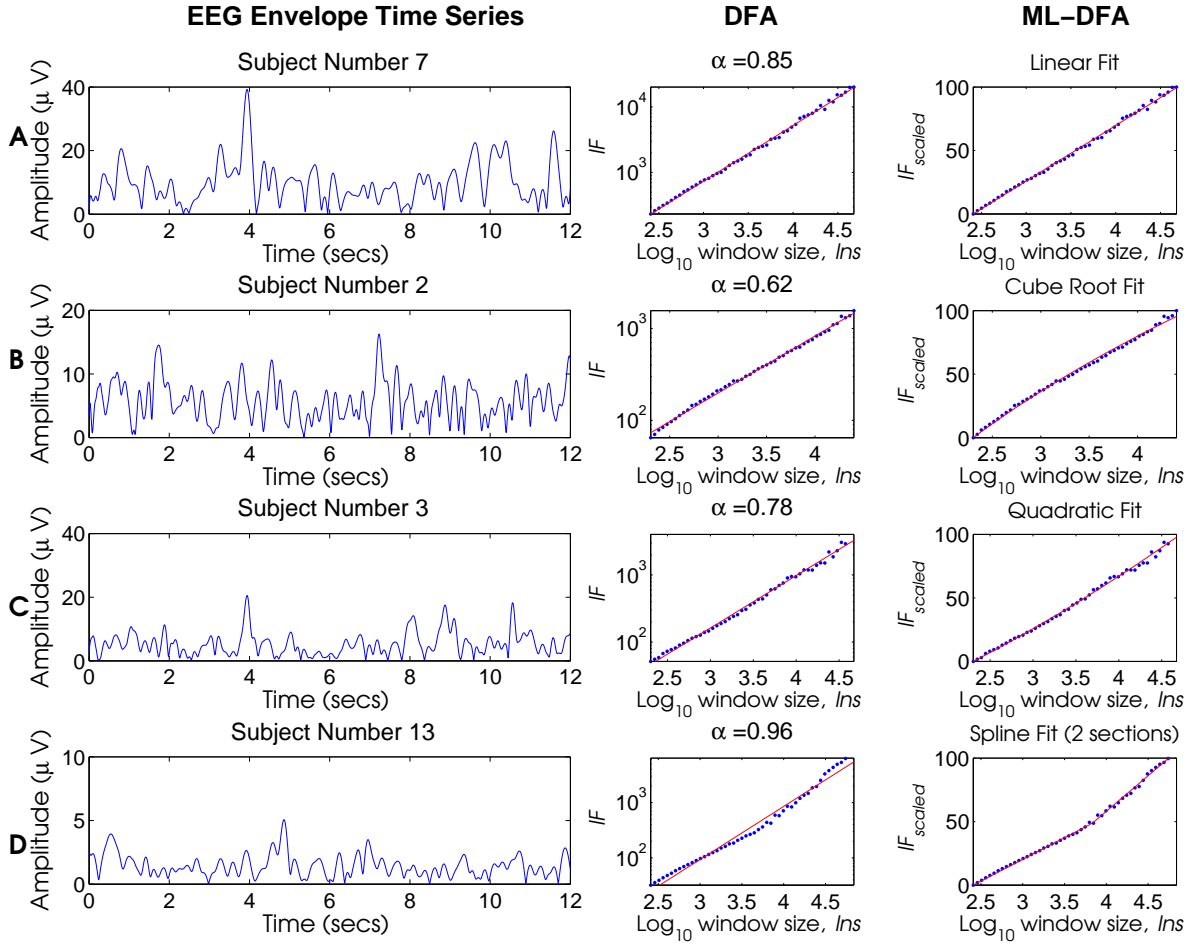}
\end{center}
\caption{ {\bf DFA fluctuation plots for 4 example signals from the Cz electrode of an EEG recording.} Specifically, the 4 rows A-D correspond to subjects 2,3,7,13. In each row, the left-hand side panel shows a representative $3000$ innovations of the time series, which corresponds to approximately 15 seconds, the middle panel shows the fluctuation plot fitted using standard DFA with the DFA exponent $\alpha$ given above each plot, and the right-hand side panel shows the best fit model as determined by ML-DFA using AIC. \newline\newline} \label{Fig6}
\end{figure}

Figure~\ref{Fig6} shows 4 examples of each of the ML-DFA fit types obtained from the 20 subjects. These data were selected to illustrate both the linear fitting by standard DFA and a range of model fits that led to the rejection of the linear model hypothesis. In total, the linear model hypothesis was not rejected in 12/20 (AIC) and 16/20 (BIC) of the subjects (see Table~\ref{healthyTab}). 

\subsubsection*{Minimum and Maximum Window Sizes}\label{sec:minmaxwinsiz}

In neurophysiological data, the choice of window sizes over which DFA is calculated is an important consideration. Using the data from one subject for which both the DFA fluctuation plot was best fit by a linear model according to both AIC and BIC, we explore how the choice of minimum and maximum window sizes affects the linearity of the DFA fluctuation plot. We demonstrate that using a minimum window length smaller than a minimum oscillatory period of the data examined gives rise to DFA fluctuation plots for which the linear model hypothesis is rejected. We also show that taking a maximum window length larger than $\frac{N}{10}$ gives rise to DFA fluctuation plots for which the linear model hypothesis may be rejected. 

\begin{figure}[!ht]
\begin{center}
\includegraphics[scale=0.5]{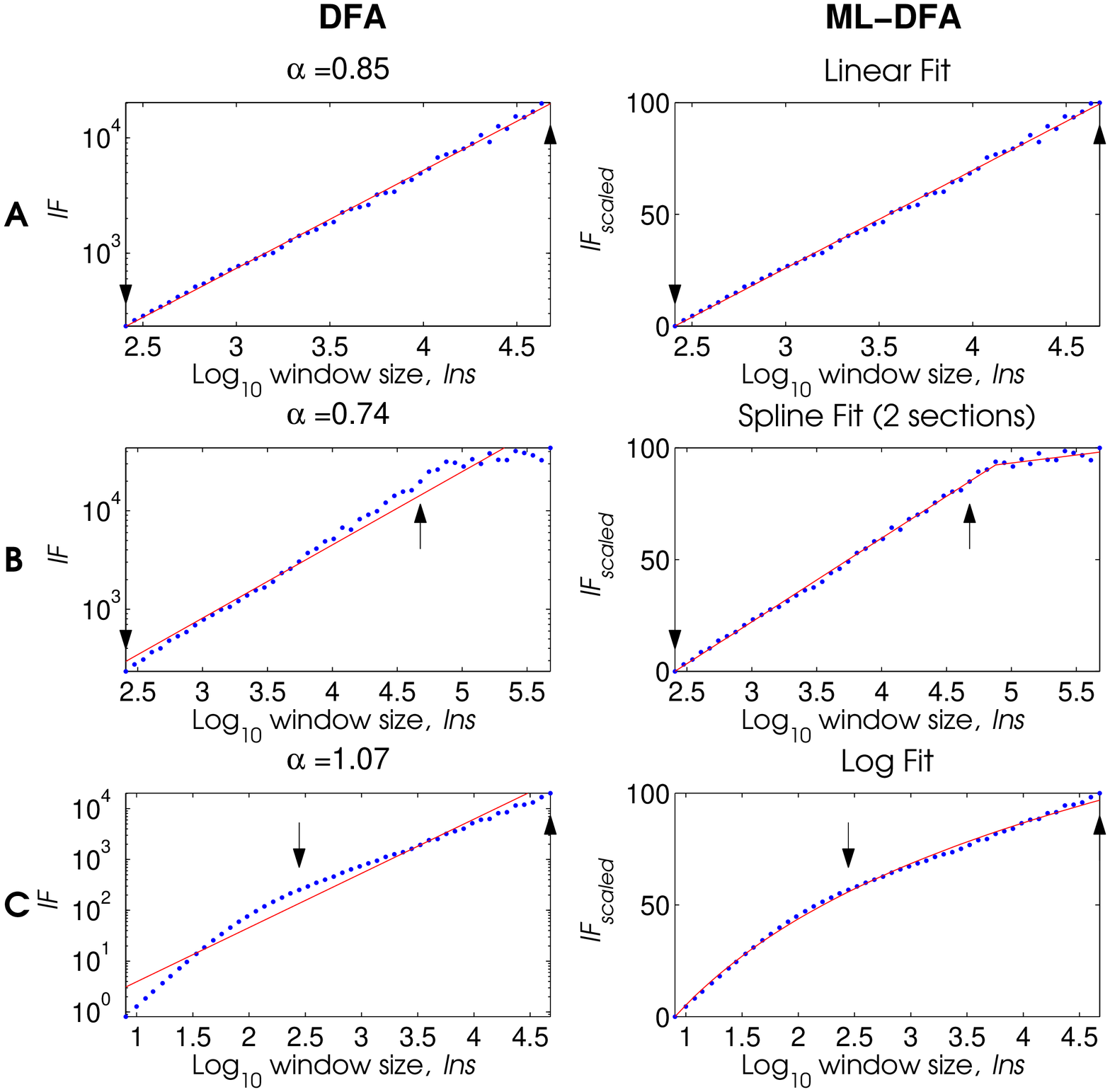}
\end{center}
\caption{ {\bf DFA fluctuation plots when different window lengths are used to analyse the Cz electrode signal of an EEG recording for subject number 7.} Panel A shows the DFA and ML-DFA analysis performed for Subject 7 in Figure~\ref{Fig6}. The minimum window size is $1$ second, and the maximum is $\frac{N}{10}$, or $187$ seconds, both following~\cite{linkenkaer01}. The arrows in each plot indicate the range over which the fluctuation plot is calculated to obtain the DFA exponents in Tables~\ref{healthyTab}, which is the full range of the fluctuation plot in Panel A. In Panel B, the minimum window length is also $1$ second of data, and the maximum window length is $N$, which is the full length of the signal, or $1865$ seconds ($31$ minutes) in this case. In Panel C, the minimum window length is $8$ samples ($\frac{8}{256} \approx 0.31$ seconds) and the maximum window length is $\frac{N}{10}$, or $187$ seconds. In each row, the left-hand side panel shows the fluctuation plots fitted using standard DFA with the DFA exponent $\alpha$ given above each plot, the right-hand side panel shows the best-fit model as determined by ML-DFA using AIC. \newline\newline} \label{Fig7}
\end{figure}

\begin{table}[ht]
\caption{\bf{Results of ML-DFA with the EEG signal obtained from the Cz electrode in 20 healthy subjects.} }  
\centering 
\begin{tabular}{|c|c|c|c|c|c|} 
\hline 
Subject Number	& Slope & AIC &	BIC	 	\\ [0.5ex] 
\hline 
1	& $0.7861^{\dagger}$&	Square Root			&	Linear			    \\
2	& $0.6204^{\dagger}$&	Cube Root			&	Cube Root			\\
3	& $0.7798^{\dagger}$&	Quadratic			&	Linear				\\
4	& $0.7504$ 			&	Linear				&	Linear				\\
5	& $0.8593^{\dagger}$&	Two-segment spline	&	Linear				\\
6	& $0.9231^{\dagger}$&	Two-segment spline	&	Two-segment spline	\\
7	& $0.8496 $			&	Linear				&	Linear				\\
8	& $0.8450$ 			&	Linear				&	Linear				\\
9	& $0.7654$ 			&	Linear				&	Linear				\\
10	& $0.7249 $			&	Linear				&	Linear				\\
11	& $0.7795$ 			&	Linear				&	Linear				\\
12	& $0.6856 $			&	Linear				&	Linear				\\
13	& $0.9595^{\dagger}$&	Two-segment spline	&	Quadratic			\\
14	& $0.9093^{\dagger}$&	Two-segment spline	&	Two-segment spline	\\
15	& $0.8762^{\dagger}$&	Two-segment spline	&	Linear				\\
16	& $0.8578$ 			&	Linear				&	Linear				\\
17	& $0.7833 $			&	Linear				&	Linear				\\
18	& $0.7631$ 			&	Linear				&	Linear				\\
19	& $0.7350$ 			&	Linear				&	Linear				\\
20	& $0.9120$ 			&	Linear				&	Linear				\\
\hline
\end{tabular}
\begin{flushleft} Each subject is identified by number in the left-hand side column, alongside the best fit model determined by ML-DFA using AIC and BIC. The $^{\dagger}$ symbol indicates those subjects whose fluctuation plots were rejected as not being linear by at least one of the AIC or BIC measures. When the fluctuation plot is rejected by BIC, it is also rejected by AIC in all cases. The exponent provided in column 2 was obtained using standard DFA. 
\end{flushleft}
\label{healthyTab} 
\end{table}

Figure~\ref{Fig7} shows the application of standard DFA to the fluctuation plots of the EEG signal subject number 7, alongside the best-fitting model determined by ML-DFA using AIC. In Panel A, the minimum and maximum window sizes are set as suggested by~\cite{linkenkaer01}. In Panel B, the minimum window length is set to $1$ second of data as previously in Figure \ref{Fig6}, while the maximum window length is $N$. The magnitude of detrended fluctuations grows more slowly for large window sizes due to the finite length of the data, and this gives rise to a two-segment spline as the best fit model such that the DFA exponent should not be trusted. In Panel C of Figure~\ref{Fig7}, the minimum window size is set to $8$ samples of the recording ($\approx 0.31$ seconds), while the maximum window length is held constant at a tenth of the length of the time series, $\frac{N}{10}$ as before. The linear model hypothesis is rejected by the AIC method, because the best-fit model is logarithmic. This is consistent with the fact that, as the signal was filtered in the $\alpha$ range of 8-13 Hz, a minimum window length less than $\frac{fs}{8}$, is less than a single cycle of the slowest $8$Hz frequency present, which will certainly produce a crossover, as shown in Figure~\ref{Fig7}. In order to select a suitable minimum window size for a signal, its characteristic frequency should be known. In this case, the characteristic frequency is not a single value, but a range between 8 and 13 Hz, so that the crossover in the fluctuation plot is not a single point (as with previously studied pure sine curves), but rather a range of points. We suggest that this is why the best fit model is the smoother logarithmic model rather than a spline. 

This analysis was applied to all $20$ subjects. When the minimum window size was taken to be $8$ samples (while keeping the maximum window size at $\frac{N}{10}$), the fluctuation plots of data for all 20 subjects were rejected as not being linear by both AIC and BIC. When the maximum window size was set to $N$, and the minimum kept at $1$ second, application of ML-DFA resulted in 4/20 (BIC) and 10/20 (AIC) fluctuation plots for which the linear model hypothesis was rejected. Once again, BIC shows less sensitivity than AIC in identifying the loss of linearity due to a strong bias towards the simplest (linear) model.  

\section*{Discussion}

In this paper, we have presented a technique (ML-DFA) to determine whether a DFA exponent can be trusted based on whether the linear model hypothesis for its associated fluctuation plot is accepted or rejected by a model selection approach. We have validated ML-DFA by applying it to DFA fluctuation plots obtained from FARIMA(0,$d$,0) time series, which have been shown to be asymptotically linear~\cite{taqqu,bardet}. We have explored ML-DFA in relation to DFA plots obtained from time series generated by FARIMA(1,d,1) processes, which allow flexible combinations of long and short correlations in the time series, and which we expect will produce fluctuation plots that are rejected as not being linear. We have recovered the piecewise linear form of the DFA fluctuation plot for sinusoidal signals, and sinusoidal signals with additive independent noise, as previously documented \cite{hu}. Finally, we applied the method to the amplitude envelopes of filtered EEG time series as in~\cite{linkenkaer01}, showing that 12 out of 20 recorded signals were not rejected by AIC and 16 out of 20 by BIC. Without the use of a test such as ML-DFA, the value of a DFA exponent could be meaningless and we suggest that it may be valuable to re-examine previously published results. 

We have stated the values of both the AIC and BIC measures throughout. It has been argued that the BIC is the most reliable information criterion \cite{daw, mackay}. However in this study, the AIC shows fewer false-positive results. This is demonstrated by the fact that a greater proportion of time series generated by FARIMA(1,$d$,1) processes are rejected as not being linear, which is a result that we would expect. Furthermore, the AIC was more successful at correctly identifying the fluctuation plots that we expected to be four-segment splines because they were obtained from a sinusoidal curve with independent, additive anti-correlated noise. The BIC often selected a quartic model instead, because it has fewer parameters. For this reason, we suggest that AIC should be used to determine the best-fitting model.

It is important to stress that ML-DFA does not verify or demonstrate the linearity of a plot. Merely, it concludes that a linear model is the best choice given the set of alternative models considered. For this reason, it is important to carefully select the set of alternative models. Since fluctuation plots should always be monotonic because the fluctuations of a time series will yield an error of at least equal size for windows of greater length, we have only considered models that (a) can capture the monotonicity of a DFA fluctuation plot and (b) are informed by experience and previous studies of non-linear DFA fluctuation plots. Note that the necessarily finite set of alternative models means that there is always a possibility that a different model could prove a better fit and therefore one should be very cautious of drawing conclusions about the nature of a time-series based on the best-fit model. Further, because the best fitting models are calculated from initial parameters that are set randomly, using a set of closely-related models with an equal number of parameters may result in different best-fit models for different runs of ML-DFA. To address this concern, initial parameters for the polynomial models were set to those that best fitted the fluctuation plots in a least square sense; however, this remains an open issue for models of arbitrary functional form. It is for these two above reasons that our focus has been primarily on whether the linear model hypothesis is rejected (which determines whether the DFA exponent can be trusted) and not on interpreting or explaining why a particular functional form was the best fitted model. 

Several papers have discussed non-linear DFA fluctuation plots for specific time series. A DFA fluctuation plot which flattens out with increasing box size typically reflects a periodic signal, such as a sine~\cite{hu}. Increasing fluctuations at large window sizes may be consistent with a noise process with segments removed, one with spikes added, one using concatenated segments of different standard deviations, or else with a power law trend~\cite{hu,chen02,chen05}. Finite-size effects cause smaller windows to always have fluctuation magnitudes below the expected regression line~\cite{bryce}. 

Additionally, a fluctuation plot can be non-linear if the DFA scaling exponent is not a single value, but comes from a distribution. In this case, it may be relevant to apply multi-fractal DFA~\cite{kantelhardt}. If the scaling behaviour of a time series is not constant across time, then a suitable technique is Adaptive Time-varying DFA~\cite{berthouze12}, which uses optimal filtering to track changes in DFA exponent over the record. Any of these considerations may help elucidate a DFA fluctuation plot for which the linear model hypothesis is rejected. 

We also varied the minimum and maximum window sizes used in the course of DFA application to highlight the fact that an inappropriate window size may affect the validity of the DFA exponent. A preliminary inspection of the whole fluctuation plot (as done by~\cite{linkenkaer01}) can be instructive for gaining a broad idea of the scales over which long-range correlations may be located. However, we stress that good practice should be to establish {\it a priori} the range of scales over which LRTCs are expected -- taking into account the constraints of both the nature of the data (e.g., sampled oscillatory data) and a statistically appropriate number of maximum window sizes -- and to accept the result returned by ML-DFA. It would be inappropriate to use this technique to identify the range of scales over which LRTCs exist. Indeed, it will always be possible to find a range of scales over which the linear model hypothesis will be accepted. 

For neurophysiological data, the minimum window size should include several oscillations of the lowest frequency, and we took $1$ second of the recording to ensure this. The frequencies present are determined by the range of the bandpass filter used. FARIMA signals do not have a characteristic time scale, so the minimum window size can be smaller, and we took $8$ innovations (a smaller window size of $1$ or $2$ innovations would have given an artefactual result because $1$ or $2$ samples can always be fitted perfectly by a line and the fluctuation magnitude will thus always be zero, a minimum window size of $4$ samples can cause inaccuracies due to finite-size effects~\cite{bryce}). For the sinusoidal curves, we also used $8$ innovations for the minimum window, which was smaller than the cycle period, precisely to allow us to demonstrate the crossovers in the fluctuation plot. The maximum window size was set to a tenth of the length of the time series for all signals considered to allow a sufficient number of values for a robust estimate of an average fluctuation size. In order to obtain a reliable fluctuation plot for larger time scales, a longer data series would typically be required~\cite{linkenkaer01}. In general, we recommend the use of these or similar guidelines for correct application of DFA and ML-DFA. Interestingly, ML-DFA makes it possible to approach the question of the maximum box size in a more systematic manner. The length of a neurophysiological time series will depend on numerous considerations, many of them experimental, and using a tenth of the data length as a maximum box size may lead to confusion when trying to infer meaning in time series of different length. Depending on the strictness of the model selection criteria between 50 and 80\% of EEG time series did not reject the linear model hypothesis even when the entire record length, i.e. $\sim$ 20 minutes, was considered. When the linear model hypothesis was rejected at large window sizes, the window size above which loss of scaling occurred could be identified (see Figure~\ref{Fig7}). We suggest therefore that ML-DFA can be used to validate relaxing a conservative choice of maximum window size (i.e., to extend the length of meaningful correlations) to help with heterogeneous lengths of time series. 

\section*{Materials and Methods}\label{sec:mm}

\subsection*{Scaling and Fit}\label{sec:fit}

DFA is used to assess the self-similarity in a signal \cite{peng94,peng95}. The application of DFA returns the value of an exponent $\alpha$, which is an estimate of the Hurst parameter, $H$, which in turn reveals the degree of long-range temporal correlation (LRTC) in the time series~\cite{hurst}. DFA can be applied to both stationary and non-stationary data, avoiding artefactual dependencies~\cite{priemer}. 

To calculate the DFA exponent, the time series is first de-meaned and then cumulatively summed. After being divided into non-overlapping windows of a given size (i.e., a scale), it is detrended (linearly for 1-DFA, non-linearly for higher-order DFA) yielding a fluctuation calculated as the root-mean-square deviation over every window at that scale. The process is repeated for different window sizes. 

For oscillatory signals, the smallest window size should be large enough to avoid errors in local root mean square fluctuations, and is typically taken to be three or four times the length of a cycle at the characteristic frequency in the time series. If the minimum window size is significantly smaller than this, then the fluctuation plot will typically contain a crossover at the window length of a single period~\cite{hu}. In the case of non-oscillatory signals such as those from a FARIMA process, there is no characteristic time scale and a smaller window size may be used. The maximum window size should be small enough to provide a robust average for the fluctuation magnitude across the time series. It is typically taken to be $N/10$ where $N$ is the length of the data~\cite{linkenkaer01}, however, a maximum window size of $N/4$ has previously also been used and shown to provide a sufficiently good estimate of the average fluctuations in some circumstances~\cite{berthouze}. 

We call $ns$ the vector of window sizes and $F$ the vector of corresponding root mean square fluctuations. We label the number of distinct window sizes $n$, which are taken as the maximum possible to allow each window to be non-overlapping. The base $10$ logarithm of these two vectors are labelled $lns$ and $lF$ respectively. 

If the signal is self-similar, then the log-log plot of fluctuation sizes against window sizes, referred to as DFA fluctuation plot throughout the manuscript, will be linear and the DFA exponent is obtained by determining the slope of the best fitting regression line. A DFA exponent in the range $0.5<\alpha<1$ indicates the presence of long-range temporal correlations. An exponent of  $0<\alpha<0.5$ is obtained when the time series is anti-correlated and $\alpha=1$ represents pink noise. Gaussian white noise has an exponent of $\alpha=0.5$. For a tutorial, see~\cite{hardstone}.    

However, since there is no {\it a priori} means of confirming that a signal is indeed self-similar, an exponent can always be obtained even though the DFA fluctuation plot may not necessarily be linear -- the only certainty being that it will be increasing (albeit not necessarily monotonously so) with window sizes. The models used by ML-DFA are listed below, with the $a_i$ parameters to be found. The number of parameters ranges between $2$ for the linear model, and $8$ for the four-segment spline model. 

\begin{list}{}{}
\item Polynomial - $f(x) = \sum_{i=0}^{K} a_i x^i \mbox{ for } K = \left\lbrace 1,...,5 \right\rbrace $
\item Root - $f(x)=a_1 (x + a_2)^{1/K} + a_3 \mbox{ for } K = \left\lbrace 2,3,4\right\rbrace $ 
\item Logarithmic - $f(x)=a_1 \mbox{log}(x + a_2) + a_3$
\item Exponential - $f(x)=a_1 e^{ a_2 x} + a_3$ 
\item Spline with 2, 3 and 4 linear sections.
\end{list}

We first normalise the fluctuation magnitudes with:
$$lF_{scaled} = 100 \times \frac{lF - lF_{min} }{lF_{max} - lF_{min}} $$
where $lF_{min}$ and $lF_{max}$ are the minimum and the maximum values of vector $lF$ respectively.
We define a likelihood function:
$$ \mathcal{L}  =  \prod_{i=1}^{n} p(lns(i))^{lF_{scaled}(i)} $$ 
which is a product across all windows $i$, where $p(lns)$ represents the function:
$$p(lns) = \frac{\vert f(lns) \vert }{\sum_{i=1}^{n} \vert f(lns) \vert }$$
where $f(lns)$ is the fitted model. Absolute values are used in order to ensure that $p(lns)$ remains in the range $\left[0, 1\right]   $, so that we reject a likelihood function if it falls below $0$. 

The log-likelihood is then defined as:
$$ \mbox{log}\mathcal{L}  =  \sum_{i=1}^{n} {lF_{scaled}(i)} \mbox{log} p(lns(i))$$   
We maximise this function to find the parameters $a_i$ necessary for $f(lns)$. 

The largest log-likelihood is the model which best fits the data, however, no consideration of the number of parameters used is taken when comparing log-likelihoods. To address this, we compute both the AIC and BIC measures which are designed to prevent over-fitting, which should in general be avoided \cite{mackay}.

It should be noted that the scaling step implies that DFA exponents cannot be recovered from the parameters of the linear or spline models following ML-DFA. For this reason, if a spline model is found to be the best-fitting model and the user is interested in the value of the exponents at each scale -- as is sometimes used in clinical studies of heart beat variability~\cite{hrvsoftware} -- then the user should apply standard DFA to each segment separately to obtain the corresponding exponents. 

\subsection*{Akaike's Information Criterion}\label{sec:AIC}
Akaike's Information Criterion (AIC) is used to compare the goodness-of-fit of probability distributions \cite{akaike}. The AIC can only be used to compare models, but gives no information on how good the model is at fitting the data. This means that only the relative values of this measure, for different models, are important. 

For a model using $k$ parameters, with likelihood function $ \mbox{log}\mathcal{L} $, the Akaike Information Criterion is calculated using the following expression:
$$ \mbox{AIC} = 2k - 2 \mbox{log}\mathcal{L} + \frac{2k(k+1)}{n-k-1}$$ 
where $k$ is the number of parameters that the model uses. Note that we are using the formula proposed by~\cite{hurvich} which accounts for small sample sizes, as advocated by~\cite{burnham,brockwell,macquarrie} amongst others. The model which provides the best fit to the data is that with the lowest value of AIC.  

\subsection*{Bayesian Information Criterion}\label{sec:BIC}

The Bayesian Information Criterion was developed by Akaike and Schwartz \cite{schwartz}. It puts harsher restrictions on the parameter number required for the model:
$$ \mbox{BIC} = - 2 \mbox{log}\mathcal{L} + k \mbox{log}(n).$$
The lowest BIC indicates the best fit model.

There is considerable discussion regarding which of the AIC or BIC measure is more effective at selecting the `correct' model, and indeed it is possible to simulate situations in which one and the other is optimal \cite{burnham}. \cite{burnham} suggests that the BIC is effective primarily when the number of observations $n$ is large enough, which may not be the case with DFA calculations with a typical number of window sizes of $50$. On the other hand, the BIC is considered more reliable because it is by construction an approximation to the Bayes factor, which is considered by many to be the only possible approach to model selection (see Chapter 1 of~\cite{daw}, and~\cite{yang} who tries to combine the two measures). 

In the analysis here, we output both the AIC and BIC measures, but ultimately base our conclusions on the AIC when the two disagree. The BIC is the stricter approach in selecting a model with the least number of parameters, however, this will lead to an undesired bias toward choosing the linear model. Our results will actually show that the AIC is more reliable in determining the best fit to fluctuation plots for signals whose functional form has previously been studied and is known.  

\subsection*{Signal Simulation}\label{sec:sigsim}

Self-similarity is a property of signals belonging to the class of signals with long-range dependence (LRD) \cite{samorodnitsky}. In order to demonstrate and test our methodology, we apply it to signals simulated using an Autoregressive Fractionally Integrated Moving Average model (FARIMA) \cite{hosking}, which provides a process that can easily be manipulated to include a variable level of short and long-term correlations within a signal, which in turn provide a broad range of DFA fluctuation plots.   

To construct a FARIMA process a sequence of zero-mean white noise is first generated, which is typically taken to be Gaussian, and necessarily so to produce fractional Gaussian noise. The FARIMA process, $X_t$, is then defined by parameters $p$, $d$ and $q$ and given by:

\begin{equation}\label{F}
\left(1-\sum_{i=1}^{p}{\phi_i B^i} \right) \left( 1-B\right)^d X_t= \left(1+\sum_{i=1}^{q}{\theta_i B^i} \right)  \varepsilon_t.
\end{equation}

$B$ is the backshift operator operator, so that $B X_t = X_{t-1}$ and $B^2 X_t = X_{t-2}$. Terms such as $(1-B)^2$ are calculated using ordinary expansion, so that $(1-B)^2 X_t = X_t - 2 X_{t-1} + X_{t-2}$. While the parameter $d$ must be an integer in the ARIMA model, the FARIMA can take fractional values for $d$. A binomial series expansion is used to calculate the result:

$$ (1-B)^d = \sum_{k=0}^{\infty} \binom{d}{k} (-B)^k.$$

The left hand sum deals with the autoregressive part of the model where $p$ indicates the number of back-shifted terms of $X_t$ to be included, $\phi_i$ are the coefficients with which these terms are weighted. The right hand sum represents the moving average part of the model. The number of terms of white noise to be included are $q$, with coefficients $\theta_i$. In the range $\vert d \vert < \frac{1}{2}$, FARIMA processes are capable of modelling long-term persistence~\cite{hosking}. As we will only consider $p=1$ and $q=1$ throughout the manuscript, we will refer to $\phi_1$ as $\phi$ and $\theta_1$ as $\theta$. We set $\vert \phi \vert < 1$, $\vert \theta \vert < 1$ to ensure that the coefficients in Equation \ref{F} decrease with increasing application of the backshift operator, thereby guaranteeing that the series converges, and $X_t$ is finite~\cite{hosking}.

A FARIMA(0,$d$,0) is equivalent to fractional Gaussian noise with $d=H-\frac{1}{2}$ \cite{hosking}. This produces a time series with a DFA fluctuation plot that has been shown to be asymptotically linear \cite{taqqu,bardet}. By manipulating the $\phi$ and $\theta$ parameters, the DFA fluctuation plots can also be distorted.

In a FARIMA(1,$d$,0) process, the $\phi$ parameter is non-zero, and an autoregressive term is added to the process. In general, an increase in the $\phi$ coefficient at constant $\theta$ induces a decrease in fluctuations for small window sizes, and a concavity is seen in the fluctuation plot \cite{morariu}. The value of $\phi$ increases the short-range exponent with an exponential relationship \cite{morariu}. This means that the process at a given time point depends linearly on the previous values in the series, so that a single impulse would affect the rest of the process infinitely far into the future. The process is expected to behave like a FARIMA(0,$d$,0) time series in the long-term, but the short term behaviour will have short-term correlations, depending on the size of $\phi$ \cite{hosking}. 

For a FARIMA(0,$d$,1) time series, the $\theta$ parameter is non-zero, which indicates exponential smoothing and a time series with noisy fluctuations around a slowly-varying mean. The resulting DFA fluctuation plots have fluctuation levels that are above the expected regression line at large box sizes. An increase in $\theta$ for $\phi=0$ induces convexity in the fluctuation plot. 

\subsection*{Neurophysiological Data}\label{sec:eeg}
A total of twenty healthy subjects were recruited from the workforce at the Royal Hospital for Neuro-disability 6 males, age range 24-59 years, of mean age 39.94 years, $\pm 10.2$. All subjects gave informed consent. Recording procedures were carried out in accordance with the declaration of Helsinki. None of the subjects had previous history of blackouts, faints, or psychiatric illness. None of the subjects were on any medication known to have centro-encephalic effects. All subjects were right handed.

The EEG recordings were conducted as part of a study exploring EEG changes occurring during music therapy.
The subjects were seated in a comfortable chair with arm rests. A total of $23$ Ag/AgCl electrodes (Unimed Electrodes, Surrey, UK) were applied individually to the scalp in accordance with the $10-20$ system of electrode placement~\cite{jasper}. Electrodes were fixed in place using Ten20 conductive paste (Weaver and Company, USA). Electrode impedances were maintained below $5K\Omega$. The EEG was acquired using an XLTEK Video-EEG monitoring system (Optima Medical, Putney, UK) which incorporated a $50$ channel amplifier. The EEG signals were acquired using a sampling rate of $256$Hz, and filter settings between $0.5-70$Hz without mains suppression. The montage regime used for on-line acquisition was common average reference~\cite{cooper}.
Recordings were taken over a period of $40-60$ minutes. The initial $5$ minutes of the recording was designated the baseline silence period (background noise $34$dB) Here, the subjects were instructed to close their eyes on hearing a series of clicks. The initial $2.5$ minutes of the baseline recording during the silence period were taken with the eyes open. Across the remainder of the session, subjects listened to different sounds/music the order of delivery having been randomly selected.

The recorded EEG signals were converted off-line to Laplacian derivation~\cite{hjorth75,hjorth80}. Artefact rejection was performed through visual inspection of the EEG and Independent Component Analysis in EEGlab~\cite{delorme}. For this reason, the length of the continuous signals subjected to analysis varied from subject to subject but had a minimum length of approximately $20$ minutes. 

\subsection*{Matlab Code}
Full code for ML-DFA is available from the corresponding author upon request. It will be made freely available upon acceptance of the manuscript. The data from FARIMA processes  were generated using Matlab code published by~\cite{stoev}.

\section*{Acknowledgments}

We would like to thank Dr C\'edric Ginestet for useful discussions, and Dr Leon James and Dr Agnieszka Kempny for allowing us to use their recordings of EEG data, which were made at the Royal Hospital of Neuro-disability, and all EEG subjects for their willing participation. Simon F. Farmer was supported by University College London Hospitals Biomedical Research Centre (BRC). Maria Botcharova thanks the Centre for Mathematics and Physics in the Life Sciences and Experimental Biology (CoMPLEX), University College London for their funding and continuing support.

\bibliography{bibliobank}

\begin{thebibliography}{10}
\providecommand{\url}[1]{\texttt{#1}}
\providecommand{\urlprefix}{URL }
\expandafter\ifx\csname urlstyle\endcsname\relax
  \providecommand{\doi}[1]{doi:\discretionary{}{}{}#1}\else
  \providecommand{\doi}{doi:\discretionary{}{}{}\begingroup
  \urlstyle{rm}\Url}\fi
\providecommand{\bibAnnoteFile}[1]{%
  \IfFileExists{#1}{\begin{quotation}\noindent\textsc{Key:} #1\\
  \textsc{Annotation:}\ \input{#1}\end{quotation}}{}}
\providecommand{\bibAnnote}[2]{%
  \begin{quotation}\noindent\textsc{Key:} #1\\
  \textsc{Annotation:}\ #2\end{quotation}}
\providecommand{\eprint}[2][]{\url{#2}}

\bibitem{peng94}
Peng CK, Buldyrev SV, Havlin S, Simons M, Stanley HE, et~al. (1994) Mosaic
  organization of {DNA} nucleotides.
\newblock Phys Rev E 49: 1685--1689.
\bibAnnoteFile{peng94}

\bibitem{peng95}
Peng CK, Havlin S, Stanley HE, Goldberger AL (1995) {Quantification of scaling
  exponents and crossover phenomena in nonstationary heartbeat time series}.
\newblock Chaos 5: 82--87.
\bibAnnoteFile{peng95}

\bibitem{hurst}
Hurst HE (1951) {Long term storage capacity in reservoirs}.
\newblock T Am Soc Civ Eng 116: 770--799.
\bibAnnoteFile{hurst}

\bibitem{clegg}
Clegg RG (2006) A practical guide to measuring the hurst parameter.
\newblock In: 21st UK Performance Engineering Workshop, School of Computing
  Science Technical Report Series, CSTR-916, University of Newcastle. pp.
  43--55.
\bibAnnoteFile{clegg}

\bibitem{granger}
Granger CWJ, Joyeux R (1980) An introduction to long-memory time series models
  and fractional differencing.
\newblock J Time Ser Anal 1: 15--29.
\bibAnnoteFile{granger}

\bibitem{varotsos}
Varotsos C, Kirk-Davidoff D (2006) Long-memory processes in ozone and
  temperature variations at the region $60^{\circ}{S} - 60^{\circ}{N}$.
\newblock Atmos Chem Phys 6: 4093--4100.
\bibAnnoteFile{varotsos}

\bibitem{robinson}
Robinson PM, editor (2003) Time series with long memory.
\newblock Oxford University Press.
\bibAnnoteFile{robinson}

\bibitem{karmeshu}
Karmeshu, Krishnamachari A (2004) Sequence variability and long-range
  dependence in {DNA}: An information theoretic perspective.
\newblock Lecture Notes in Computer Science 3316: 1354-1361.
\bibAnnoteFile{karmeshu}

\bibitem{karagiannis}
Karagiannis T, Molle M, Faloutsos M (2004) {Long-range dependence: Ten years of
  Internet traffic modeling}.
\newblock IEEE Internet Comput 8: 57--64.
\bibAnnoteFile{karagiannis}

\bibitem{samorodnitsky}
Samorodnitsky G (2006) Long range dependence.
\newblock Found Trends Stoch Syst 1: 163--257.
\bibAnnoteFile{samorodnitsky}

\bibitem{chialvo}
Chialvo DR (2010) {Emergent complex neural dynamics}.
\newblock Nature Physics 6: 744--750.
\bibAnnoteFile{chialvo}

\bibitem{sornette}
Sornette D (2006) {Critical Phenomena in Natural Sciences: Chaos, Fractals,
  Selforganization and Disorder: Concepts and Tools}.
\newblock Springer, 2nd edition.
\bibAnnoteFile{sornette}

\bibitem{timme}
Beggs JM, Timme N (2012) Being critical of criticality in the brain.
\newblock Front Fract Physiol 3: 163.
\bibAnnoteFile{timme}

\bibitem{stam}
Stam CJ, de~Bruin EA (2004) {Scale-free dynamics of global functional
  connectivity in the human brain}.
\newblock Hum Brain Mapp 22: 97--109.
\bibAnnoteFile{stam}

\bibitem{werner}
Werner G (2010) {Fractals in the nervous system: conceptual implications for
  theoretical neuroscience.}
\newblock Front Physiol 1: 15.
\bibAnnoteFile{werner}

\bibitem{linkenkaer01}
Linkenkaer-Hansen K, Nikouline VV, Palva JM, Ilmoniemi RJ (2001) Long-range
  temporal correlations and scaling behavior in human brain oscillations.
\newblock J Neurosci 21: 1370-7.
\bibAnnoteFile{linkenkaer01}

\bibitem{linkenkaer04}
Linkenkaer-Hansen K, Nikulin VV, Palva JM, Kaila K, Ilmoniemi RJ (2004)
  Stimulus-induced change in long-range temporal correlations and scaling
  behaviour of sensorimotor oscillations.
\newblock Eur J Neurosci 19: 203--218.
\bibAnnoteFile{linkenkaer04}

\bibitem{maraun04}
Maraun D, Rust HW, Timmer J (2004) {Tempting long-memory - on the
  interpretation of DFA results}.
\newblock Nonlinear Proc Geoph 11: 495--503.
\bibAnnoteFile{maraun04}

\bibitem{hu}
Hu K, Ivanov PC, Chen Z, Carpena P, Eugene~Stanley H (2001) Effect of trends on
  detrended fluctuation analysis.
\newblock Phys Rev E 64: 011114.
\bibAnnoteFile{hu}

\bibitem{chen02}
Chen Z, Ivanov P, Hu K, Stanley HE (2002) Effect of nonstationarities on
  detrended fluctuation analysis.
\newblock Phys Rev E 65: 041107.
\bibAnnoteFile{chen02}

\bibitem{chen05}
Chen Z, Hu K, Carpena P, Bernaola-Galvan P, Stanley HE, et~al. (2005) Effect of
  nonlinear filters on detrended fluctuation analysis.
\newblock Phys Rev E 71: 011104.
\bibAnnoteFile{chen05}

\bibitem{grech}
Grech D, Mazur Z (2013) Scaling range of power laws that originate from
  fluctuation analysis.
\newblock Phys Rev E 87: 052809.
\bibAnnoteFile{grech}

\bibitem{anscombe}
Anscombe FJ (1973) {Graphs in Statistical Analysis}.
\newblock Amer Statist 27: 17--21.
\bibAnnoteFile{anscombe}

\bibitem{raymond}
Raymond GM, Bassingthwaighte JB (1999) Deriving dispersional and scaled
  windowed variance analyses using the correlation function of discrete
  fractional gaussian noise.
\newblock Physica A 265: 85--96.
\bibAnnoteFile{raymond}

\bibitem{bardet}
Bardet JM, Kammoun I (2008) Asymptotic properties of the detrended fluctuation
  analysis of long-range-dependent processes.
\newblock IEEE T Inform Theory 54: 2041--2052.
\bibAnnoteFile{bardet}

\bibitem{akaike}
Akaike H (1974) {A new look at the statistical model identification}.
\newblock IEEE T Automat Contr 19: 716--723.
\bibAnnoteFile{akaike}

\bibitem{schwartz}
Schwarz G (1978) {Estimating the Dimension of a Model}.
\newblock Ann Statist 6: 461--464.
\bibAnnoteFile{schwartz}

\bibitem{hosking}
Hosking JRM (1981) {Fractional differencing}.
\newblock Biometrika 68: 165--176.
\bibAnnoteFile{hosking}

\bibitem{taqqu}
Taqqu MS, Teverovsky V, Willinger W (1995) Estimators for long-range
  dependence: An empirical study.
\newblock Fractals 3: 785--798.
\bibAnnoteFile{taqqu}

\bibitem{mcewen}
McEwen J, Anderson G (1975) Modeling the stationarity and gaussianity of
  spontaneous electroencephalographic activity.
\newblock IEEE Trans Biomed Eng 22: 361--9.
\bibAnnoteFile{mcewen}

\bibitem{runstest}
Wald A, Wolfowitz J (1940) {On a test whether two samples are from the same
  population}.
\newblock Ann Math Statist 11: 147--162.
\bibAnnoteFile{runstest}

\bibitem{daw}
Delgado MR, Phelps EA, Robbins TW, editors (2011) Decision Making, Affect, and
  Learning: Attention and Performance XXIII.
\newblock Oxford University Press.
\bibAnnoteFile{daw}

\bibitem{mackay}
Mackay DJC (2003) {Information Theory, Inference and Learning Algorithms}.
\newblock Cambridge University Press, first edition.
\bibAnnoteFile{mackay}

\bibitem{bryce}
Bryce RM, Sprague KB (2012) {Revisiting detrended fluctuation analysis}.
\newblock Sci Rep 2: 315.
\bibAnnoteFile{bryce}

\bibitem{kantelhardt}
Kantelhardt JW, Zschiegner SA, Bunde EK, Havlin S, Bunde A, et~al. (2002)
  {Multifractal detrended fluctuation analysis of nonstationary time series}.
\newblock Physica A 316: 87--114.
\bibAnnoteFile{kantelhardt}

\bibitem{berthouze12}
Berthouze L, Farmer SF (2012) Adaptive time-varying detrended fluctuation
  analysis.
\newblock J Neurosci Methods 209: 178--188.
\bibAnnoteFile{berthouze12}

\bibitem{priemer}
Priemer R (1990) Introductory Signal Processing.
\newblock River Edge, NJ, USA: World Scientific Publishing.
\bibAnnoteFile{priemer}

\bibitem{berthouze}
Berthouze L, James LM, Farmer SF (2010) {Human EEG shows long-range temporal
  correlations of oscillation amplitude in Theta, Alpha and Beta bands across a
  wide age range}.
\newblock Clin Neurophysiol 121: 1187--1197.
\bibAnnoteFile{berthouze}

\bibitem{hardstone}
Hardstone R, Poil SS, Schiavone G, Jansen R, Nikulin VV, et~al. (2012)
  Detrended fluctuation analysis: A scale-free view on neuronal oscillations.
\newblock Front Physiol 3: 450.
\bibAnnoteFile{hardstone}

\bibitem{hrvsoftware}
Tarvainen M, Niskanen JP, Lipponen J, Ranta-aho P, Karjalainen P (2009) Kubios
  {HRV} -- {A} software for advanced heart rate variability analysis.
\newblock In: Sloten J, Verdonck P, Nyssen M, Haueisen J, editors, 4th European
  Conference of the International Federation for Medical and Biological
  Engineering, Springer Berlin Heidelberg, volume~22 of \emph{IFMBE
  Proceedings}. pp. 1022-1025.
\bibAnnoteFile{hrvsoftware}

\bibitem{hurvich}
Hurvich CM, Tsai CL (1989) {Regression and time series model selection in small
  samples}.
\newblock Biometrika 76: 297--307.
\bibAnnoteFile{hurvich}

\bibitem{burnham}
Burnham K, Anderson D (2002) {Model selection and multimodel inference: A
  practical information-theoretic approach}.
\newblock Springer, 2nd edition.
\bibAnnoteFile{burnham}

\bibitem{brockwell}
Brockwell PJ, Davis RA (1991) {Time series: Theory and methods}.
\newblock Springer-Verlag.
\bibAnnoteFile{brockwell}

\bibitem{macquarrie}
MacQuarrie A, Tsai C (1998) Regression and time series model selection.
\newblock World Scientific Publishing.
\bibAnnoteFile{macquarrie}

\bibitem{yang}
Yang Y (2005) Can the strengths of {AIC} and {BIC} be shared? {A} conflict
  between model identification and regression estimation.
\newblock Biometrika 92: 937--950.
\bibAnnoteFile{yang}

\bibitem{morariu}
Morariu VV, Buimaga-Iarinca L, Vamos C, Soltuz S (2007) Detrended fluctuation
  analysis of autoregressive processes.
\newblock arXiv:07071437 .
\bibAnnoteFile{morariu}

\bibitem{jasper}
Jasper HH (1958) {The ten-twenty electrode system of the International
  Federation}.
\newblock Electroencephalogr Clin Neurophysiol 10: 371--375.
\bibAnnoteFile{jasper}

\bibitem{cooper}
Cooper R, Osselton J, Shaw J (1974) EEG technology.
\newblock Butterworth-Heinemann Limited.
\bibAnnoteFile{cooper}

\bibitem{hjorth75}
Hjorth B (1975) {An on-line transformation of EEG scalp potentials into
  orthogonal source derivations.}
\newblock Electroencephalogr Clin Neurophysiol 39: 526--530.
\bibAnnoteFile{hjorth75}

\bibitem{hjorth80}
Hjorth B (1980) {Source derivation simplifies topographical EEG
  interpretation}.
\newblock Am J EEG Technol 20: 121--132.
\bibAnnoteFile{hjorth80}

\bibitem{delorme}
Delorme A, Makeig S (2004) {EEGLAB: An Open Source Toolbox for Analysis of
  Single-Trial EEG Dynamics Including Independent Component Analysis}.
\newblock J Neurosci Methods 134: 9--21.
\bibAnnoteFile{delorme}

\bibitem{stoev}
Stoev S, Taqqu MS (2004) Simulation methods for linear fractional stable motion
  and {FARIMA} using the fast {F}ourier transform.
\newblock Fractals 12: 95--121.
\bibAnnoteFile{stoev}

\end{thebibliography}



\end{document}